\documentclass[prd,preprintnumbers,final]{revtex4}
 \usepackage{graphicx}
  \usepackage{bm}
   \usepackage{amsmath,amsthm}
    \usepackage{amssymb}
     \usepackage{pifont}
      \usepackage{srcltx}

\newcommand{\nn}{\nonumber}

\newcommand{\tr}{\mathrm{tr}}

\renewcommand{\(}{\left(}
\renewcommand{\)}{\right)}
\renewcommand{\[}{\left[}
\renewcommand{\]}{\right]}

\newcommand{\aboth}[1]{\stackrel{\leftrightarrow}{#1}}

\begin{document}
\title{TMD PDFs in the Laguerre polynomial basis}
\author{~A.A.Vladimirov}
\affiliation{ Department of Astronomy and Theoretical Physics, Lund University,\\ \small S\"olvegatan 14A, S 223 62 Lund, Sweden}
\email{vladimirov.aleksey@gmail.com}

\preprint{LU TP 14-04}\preprint{ Feb. 2014}

\begin{abstract}
We suggest the modified matching procedure for TMD PDF to the integrated PDF aimed to increase the amount of perturbative information in the TMD
PDF expression. The procedure consists in the selection and usage of the non-minimal operator basis, which restricts the expansion to desired
general behavior. The implication of OPE allows to systematic account of the higher order corrections. In the case of TMD PDF we assume the
Gaussian behavior, which suggests Laguerre polynomial basis as the best for the convergence of OPE. We present the leading and next-to-leading
expression of TMD PDF in this basis. The obtained perturbative expression for the TMD PDF is valid in the wide region of $b_T$ (we estimate this
region as $b_T\lesssim 2-3$ GeV$^{-1}$ depending on $x$).
\end{abstract}
\maketitle

\section{Introduction}

Transverse momentum dependent (TMD) parton distribution functions (PDFs) and fragmentation functions (FFs) (we will refer them collectively as
TMDs) contain a mixture of both perturbative and non-perturbative contribution. There are no strick rules and prescriptions for the separation
of these contribution. Depending on the details on the definition one can add or subtract some parts of perturbative expansion to a
non-perturbative contributions. In this paper we suggest new point of view on the perturbative contribution to TMDs. The main goal of our
reformulation is to increase the role of the perturbative input in the description of the TMDs.

The TMD factorization theorems give the relation between the TMDs and transversely differential cross-sections in the limit $Q^2\gg
b_T^{-2},\Lambda^2$, where $Q^2$ is the large external momentum, $b_T$ is the Fourier conjugated variable  to the relative transverse momentum
of hadrons, and $\Lambda$ is a typical intrinsic hadronic scale. As a result of factorization procedure, TMDs depend on $x$ (the longitudinal
part of the parton momentum), $b_T$, and scales of the factorization procedure $\mu$ and $\zeta$. While the dependence on the scales $\mu$ and
$\zeta$ is known, via the corresponding evolution equations, the $x$- and $b_T$-dependence are the subject of fitting and modeling. For a
theoretical introduction to the TMD factorization theorems and TMDs see \cite{Collins:2011zzd,Collins:1989gx}, while for the recent
phenomenological review see e.g.\cite{Boer:2011fh,Aidala:2014hva,Echevarria:2014xaa} and references therein.

In order to get in touch with the integrated parton distribution function, as well as to increase the  perturbative QCD input the additional
factorization procedure is applied \cite{Collins:1981uk}. This procedure is based on the operator product expansion (OPE) at short transverse
distances (we will refer it simply as OPE), and it is well-founded for $b_T^{-2}\gg \Lambda^2$. At extremely small-$b_T$ the expansion is
saturated by the first term of OPE, which is proportional to the integrated parton distribution. For larger $b_T$ the higher terms of the OPE
should be taken into account. These terms result to some unknown parton distributions.

Usually, the contributions of higher order terms of OPE are replaced by a single unknown function, which at $b_T\to 0$ reduces to unity, see
e.g. \cite{Collins:1981va}. This function is called non-perturbative factor. For the details of the non-perturbative factor introduction and its
modern status see \cite{Aidala:2014hva,Aybat:2011zv,Aybat:2011ge,Bacchetta:2013pqa}. Within the perturbative QCD the non-perturbative factor
cannot be obtained analytically, but only extracted from the comparison with data. Frequently, the non-perturbative factor is taken in the form
of a Gaussian exponent see e.g.\cite{Aidala:2014hva,Echevarria:2014xaa,Aybat:2011zv,Bacchetta:2013pqa}, so we may conclude that the contribution
of the higher OPE terms is significant. At the same time, the details of the fine structure and the intrinsic scale dependence of
non-perturbative factor are unknown and vary between studies, see e.g. discussion in \cite{Aidala:2014hva}, and references therein.

In order to proceed further we should clarify the notions of perturbative and non-perturbative contributions. Generally speaking, within QCD one
can perform the small-$b_T$ OPE of the TMD operator up to any given $\alpha_s$- and operator-order. Hence, this part of information, namely, the
coefficient functions of operators, is entirely perturbative. The truly non-perturbative objects are the hadronic states. Therefore, taking the
matrix element of the small-$b_T$ OPE we obtain products of perturbative coefficient functions with non-perturbative parton distributions. In
such a way, the non-perturbative factor contains a mixture of perturbative and non-perturbative parts. The important question is how to maximize
the perturbative input with the minimum number of non-perturbative functions. In the following we suggest a reorganization of the small-$b_T$
expansion, which may increase the amount of perturbative information in TMDs.

The standard small-$b_T$ OPE is ordered by the powers of $b_T$. Symbolically, it can be written as
\begin{eqnarray}\label{intro:O=G_nO_n}
O(x,b_T)=\sum_{n=0}^\infty G^{(T)}_n(x,b_T)\otimes O^{(T)}_n(x),
\end{eqnarray}
where $O(x,b_T)$ is a TMD operator, $O_n$ is a transversally local operator (operator with only light-cone nonlocality), and $\otimes$
represents the Mellin convolution with respect to $x$. The coefficient function $G_n$ is proportional to $b_T^n$. In the absence of interaction
the right-hand-side of (\ref{intro:O=G_nO_n}) represents the Taylor series of the operator $O(x,b_T)$ at $b_T=0$ (therefore, we mark the
operators and coefficient function in (\ref{intro:O=G_nO_n}) with superscript T). In this context, the operators $O_n$ are proportional to
$n$'th power of transverse derivative, $O_n\sim
\partial^n$.

The Taylor-like OPE is well-founded in the presence of an extreme parameter, e.g. at small $b_T$, or large $k_T$. In these cases the
contribution of the higher terms of OPE is under control. In the absence of the extreme parameter, the contribution of the higher OPE terms is
uncontrolled.

In the case of middle $b_T$ the Taylor-like OPE cannot guaranty the smallness of higher terms. Moreover, it is well-known that at the middle
$b_T$ TMDs have rapid behavior, which cannot be described by several first terms of a Taylor-like expansion. Altogether it shows that the
Taylor-like OPE is inefficient for the description of TMDs. And probably, the OPE performed in some other basis of operators, would show better
convergence at middle-$b_T$.

The operator basis for the small-$b_T$ expansion should satisfy several general assumptions. First of all, the operator basis should be
transversally local. Second, the operator basis should be orthogonal, at least in the free theory. This demand is necessary for the universal
definitions of the parton distributions. Third, the operators should be defined on the two dimensional plane for the general case, and on the
ray from the zero to infinity for the unpolarized case. Additionally, one can impose symmetry or other constraints, which follow from the
auxiliary guidelines.

The first two assumptions imply that the basis for the OPE should be chosen in the class of orthogonal polynomials. The third assumption gives
some constraints on these set of operators, but do not fix the basis unambiguously. In the absence of the external assumptions we limit ourself
to the classical orthogonal polynomials. Within classical orthogonal polynomials there are only two types of polynomials which satisfy the
demands \cite{Bateman2}.

The first one is the Hermite polynomials $H_n(x)$. They are orthogonal on the range $x\in (-\infty,+\infty)$, and therefore, suites for the
expansion of some general TMD operator
\begin{eqnarray}\label{intro:O=G*H}
O(x,b_T)=\sum_{n=0}^\infty G^{(H)}_{n_1n_2}(x,b_T)\otimes O^{(H)}_{n_1n_2}(x),
\end{eqnarray}
where $O^{(H)}_{n_1n_2}\sim H_{n_1}(\partial_1)H_{n_2}(\partial_2)$ with subscript 1 and 2 denoting the components of the transverse vector.
Such choice of the operators is not very convenient.

The second one is the Laguerre polynomials $L_n(x)$. They are orthogonal on the range $x\in (0,\infty)$. Therefore, they suite for the OPE of
the direction independent TMD operators
\begin{eqnarray}\label{intro:O=G*L}
O(x,|b_T|)=\sum_{n=0}^\infty G^{(L)}_{n}(x,b_T)\otimes O^{(L)}_{n}(x),
\end{eqnarray}
where $O^{(L)}_n\sim L_{n}(\partial^2)$. In particular, due to the simple relations between Laguerre polynomials and Hermite polynomials there
is a natural relation between expansion (\ref{intro:O=G*H}) and (\ref{intro:O=G*L}). Namely, averaging (\ref{intro:O=G*H}) over angles one
obtain (\ref{intro:O=G*L}). In this article we consider only the unpolarized TMD PDF, and therefore, we are concentrated on the Laguerre
polynomials only. For brevity, we will call the OPE with the operators in the form the Laguerre polynomials, as Laguerre-based OPE.

We want to stress that there is no definite choice of OPE basis. Our choice of the Laguerre and Hermite polynomials is based only on their
simplicity and common knowledge of the these polynomials. Nonetheless, it is not a bad choice. In particular, the usage of Laguerre or Hermite
polynomials for OPE guaranties the dominating Gaussian behavior of coefficient functions at middle and large $b_T$. This is because the
generation function for the Laguerre and Hermite polynomials have Gaussian behavior. The Gaussian anzatz is often used for fitting TMDs and it
describes $b_T$-dependence well enough. Therefore, we can expect that the OPE over operators in form of Laguerre polynomials would saturate
experimental data by less terms comparing to the Taylor-like expansion.

In the both cases, Taylor-like expansion (\ref{intro:O=G_nO_n}) and Laguerre-based expansion (\ref{intro:O=G*L}), one faces many operators of
higher orders, which matrix elements are unknown. Only the leading term in both expansions results to a known distribution, namely, integrated
PDF. Therefore, for the practical application one uses only the first term of OPE. Thus, the Laguerre-based version of OPE does not eliminate
the necessity to introduce the non-perturbative factor. However, probably, the new version of OPE reduces the significance of the
non-perturbative factor, and increases the amount of calculable input.

The most important part of OPE consideration is the loop-corrections to the coefficient function. The corrections give the deviation of the
functional form from the free-theory limit. In the Taylor-like OPE the corrections can contain only the logarithms of $b_T$, due to the
renormalizability of the theory. In the non-Taylor-like expansion, the other type of corrections are possible, e.g. power corrections. These
corrections are of special interest, because they produce the perturbative deviation from the Gaussianity. The observation of such a deviation
in the experimental data can justify the usage of the Laguerre-based approach.

In the paper we derive the coefficient function for leading term of the Laguerre-based OPE of the unpolarized TMD PDF operator, and discuss its
properties. The paper is structured as follow. In order to introduce the necessary algebraic relation and to discuss the technical details of
the Laguerre-based OPE, in sec.I we consider TMD operator in the toy-example of $\phi^3$-theory in six dimensions.  Then in sec.II we present
calculation of the coefficient function for the Laguerre-based OPE in QCD. The discussion of the behavior of TMD PDF in the Laguerre-based OPE
is presented in sec.II.2.

\section{Small-$b_T$ expansion in $\phi^3$ theory}

In order to demonstrate the details of the Laguerre-based OPE without excessive technicalities we start from the consideration of OPE in the
$\phi^3$ theory in six dimensions. For shortness we refer this model as $\phi^3_{D=6}$-model. It has similar to QCD structure of diagrams and
with a slight effort of imagination it can be transformed to QCD almost in everything, except gauge links. The action of the model reads
\begin{eqnarray}\label{phi3:action}
S=\int d^6 x \(-\frac{1}{2}\phi\partial^2\phi-\frac{g}{3!}\phi^3\).
\end{eqnarray}
An analog of QCD TMD PDF operator is
\begin{eqnarray}\label{phi3:Op}
O(x,b_T,\mu)=p_+Z_\phi(\mu)\int \frac{d\xi^-}{2\pi} e^{-ixp_+ \xi^-} \phi_r\(\xi/2\)\phi_r\(-\xi/2\),
\end{eqnarray}
where $\xi$ denotes an arbitrary light-front vector with components $\{0,\xi^-,b_T\}$. The coefficient $Z_\phi$ denotes the renormalization
constant for the field $\phi$, and $\phi_r$ denotes the renormalized field $\phi$.

At small $b_T$ the OPE of (\ref{phi3:Op}) has the form
\begin{eqnarray}\label{phi3:OPE_simple}
&&O(x,b_T)=
\\\nn&&\int_x^1 \frac{dz}{z}C_0(z,b_T^2\kappa^2,\mu^2)O_0\(\frac{x}{z},\kappa^2\)+ \int_x^1 \frac{dz}{z}b^\mu_T
C_1(z,b_T^2\kappa^2,\mu^2)O^\mu_1\(\frac{x}{z},\kappa^2\)+...,
\end{eqnarray}
where
\begin{eqnarray}\label{phi3:op_examples}
O_0\(\frac{x}{z},\kappa^2\)&=&p_+Z_\phi(\kappa)\int\frac{d\xi^-}{2\pi} e^{-ixp_+ \xi^-} \phi_r\(\xi^-/2\)\phi_r\(-\xi^-/2\),\\\nn
O_1^\mu\(\frac{x}{z},\kappa^2\)&=&p_+Z_\phi(\kappa)\int\frac{d\xi^-}{2\pi} e^{-ixp_+ \xi^-} \phi_r\(\xi^-/2\)\aboth{\partial^\mu_\perp}
\phi_r\(-\xi^-/2\),
\end{eqnarray}
and the dots stay for the terms with operators of the higher dimensions. The operators of the higher dimensions include the operators with
larger number of derivatives, as well as operators with larger number of fields. The coefficient functions $C_n$ are dimensionless, they are
functions of $x$, $g(\mu)$ and the logarithms of $b_T\kappa$.

The scale $\kappa$ in the OPE (\ref{phi3:OPE_simple}) is the scale of the subtractions of the operator singularities. This scale is independent
on the renormalization scale of the theory, which is denoted as $\mu$. However, usually it is simpler to set $\kappa=\mu$. In particular, in the
operators (\ref{phi3:op_examples}) we suppose the renormalization of the field $\phi$ at the scale $\kappa$. The difference between the
normalization points is placed into the coefficient function. Such a composition of normalization points can be always achieved with the help of
corresponding evolution equations, if the difference between $\kappa$ and $\mu$ is small enough.

Our aim is to redefine the operators of OPE (\ref{phi3:OPE_simple}) in such a way, that the dominating behavior of coefficient functions would
be Gaussian. This implies the reorganization of the whole OPE. However, since the matrix elements of the higher terms of OPE are unknown (in
contrast to the matrix element of operator $O_0$, which is the integrated PDF), our main aim is the leading term of reorganized OPE. In
particular, it means that the terms with four and higher number of fields are uninteresting to us. At the same time, we have to consider
operators with arbitrary number of derivatives in the sector of two-field operators, because the desired reorganization involves coefficient
functions of all such operators.

\subsection{OPE in the free theory}

To start with, let us consider the OPE in the free theory. In the free theory the fields can be interpreted as a classical fields and the OPE at
small $b_T$ is just a Taylor expansion at $b_T=0$. Thus, the expression (\ref{phi3:OPE_simple}) in the free-theory reads
\begin{eqnarray}\label{phi3:OP_Taylor_free_all}
&&O(x,b_T)= \\\nn && p_+\sum_{n=0}^\infty\int_x^1 \frac{dz}{z}\frac{b_T^{\mu_1}...b_T^{\mu_n}}{n!}\delta\(1-\frac{x}{z}\)
\int\frac{d\xi^-}{2\pi} e^{-izp_+ \xi^-} \phi\(\xi^-/2\)\aboth{\partial^{\mu_1}_T}...\aboth{\partial^{\mu_n}_T}\phi\(-\xi^-/2\).
\end{eqnarray}
Here we omit the field-renormalization constants and the field labels. The subscript $T$ on the vectors and scalar products denotes the
transverse component of the vectors and Euclidian scalar product.

In the following we are interesting only in the unpolarized TMDs. Therefore, we neglect the operators, which do not contribute to the
unpolarized case. It can be archived by averaging over the directions of $b_T$. Averaging both sides of (\ref{phi3:OP_Taylor_free_all}) we
obtain
\begin{eqnarray}\label{phi3:OPE_Taylor_free}
O(x,|b_T|)&=&p_+\sum_{n=0,2,..}^\infty \frac{(b_T^2)^\frac{n}{2}}{2^n
\(\frac{n}{2}\)!\(\frac{d_\perp}{2}\)_{\frac{n}{2}}}\int\frac{d\xi^-}{2\pi} e^{-ixp_+ \xi^-}
\phi\(\xi^-/2\)(\aboth{\partial^{2}_T})^{\frac{n}{2}}\phi\(-\xi^-/2\),
\end{eqnarray}
where $d_\perp=4$ is the dimension of the transverse space. We will keep $d_\perp$ arbitrary for the $\phi^3_{D=6}$ model in order to have clear
connection with the QCD case considered in the sec.3.

In order to rewrite the Taylor-like OPE via the Laguerre polynomial basis we use the following relation
\begin{eqnarray}\label{phi3:x^n->L_n(x)}&&
\phi\(\xi^-/2\)(\aboth{\partial^{2}_T})^{\frac{n}{2}}\phi\(-\xi^-/2\)=\\&&\nn(B_T^2)^{-\frac{n}{2}}\sum_{k=0,2,..}^n (-1)^\frac{n+k}{2}
\frac{\(\frac{n+d_\perp-2}{2}\)!\(\frac{n}{2}\)!}{\(\frac{k+d_\perp-2}{2}\)!\(\frac{n-k}{2}\)!}\phi\(\xi^-/2\)
L_{\frac{k}{2}}^\frac{d_\perp-2}{2}\(-\aboth{\partial^2_T}B_T^2\)\phi\(-\xi^-/2\),
\end{eqnarray}
where $L^{\frac{d_\perp-2}{2}}_n(x)$ are the associated Laguerre polynomials of order $n$ and degree $\frac{d_\perp-2}{2}$. The scale $B_T$ is
some constant scale which needed for the compensation of the dimension in the argument of the Laguerre polynomial. In fact, the right-hand-side
of (\ref{phi3:x^n->L_n(x)}) is independent on $B_T$.

The expression (\ref{phi3:x^n->L_n(x)}) is an algebraic expression, and therefore, it is valid for the interacting fields, as well as for the
free theory. With the help of this expression one can rewrite the two-field sector of any Teylor-like OPE via the Laguerre-based operators.
Inserting the expansion (\ref{phi3:x^n->L_n(x)}) into (\ref{phi3:OPE_Taylor_free}) we obtain
\begin{eqnarray}\label{phi3:OPE_Laguerre_free}
O(x,|b_T|)&=&\sum_{n=0,2,..}^\infty \(\frac{b_T^2}{4B_T^2}\)^\frac{n}{2}e^{-\frac{b_T^2}{4B_T^2}}\mathbb{O}_n\(x,B_T\),
\end{eqnarray}
where operator $\mathbb{O}$ reads
\begin{eqnarray}\label{phi3:Op_Laguerre}
\mathbb{O}_n(x,B_T)=p_+\frac{\(\frac{d_\perp-2}{2}\)!}{\(\frac{n+d_\perp-2}{2}\)!}\int\frac{d\xi^-}{2\pi} e^{-ixp_+ \xi^-}
\phi\(\xi^-/2\)L^\frac{d_\perp-2}{2}_\frac{n}{2}\(-\aboth{\partial^{2}_T}B_T^2\)\phi\(-\xi^-/2\)\!\!.
\end{eqnarray}

One can see that, indeed, the dominating behavior of the coefficient functions in Laguerre-based OPE is Gaussian. Therefore, we can expect that
such an expansion would saturate faster, comparing with the expansion (\ref{phi3:OPE_Taylor_free}). Such an expectation cannot be proved by any
perturbative calculation due to the lack of extreme parameters. The only reason for our hope is the experimental observation that the dominating
behavior of the TMDs is Gaussian.

The expression (\ref{phi3:OPE_Laguerre_free}) is independent on the parameter $B_T$. This independence is of the algebraic origin and,
therefore, one can derive the relation between the operators with different $B_T$. However, such relation involves operators of different
orders, and therefore, it is useless in the absence of information about the higher order terms.

The original OPE can be obtained by reexpanding (\ref{phi3:OPE_Laguerre_free}) at $b_T=0$ or by taking the limit $B_T\to+\infty$. Lowering the
parameter $B_T$ one takes away the parts from the lower terms of OPE and distributes them between higher terms. In this way, one may think of
the variation over $B_T$, as about redistribution of the coefficient functions between the operators.

\subsection{Laguerre-based OPE at one loop}

The consideration of the OPE in the free theory is almost an algebraic exercise. The more interesting subject is consideration of the
perturbative corrections to the coefficient functions. We will follow the strategy of the previous paragraph: firstly, we derive the coefficient
functions for the Taylor-like expansion at arbitrary order; secondly, using the relation (\ref{phi3:x^n->L_n(x)}) we reorganize the series via
the operators in the form of Laguerre polynomials. In principal, the perturbative corrections should not be of the same $b_T$-behavior as the
leading term. This is because the coefficient functions of  the non-Taylor-like OPE can contain non-logarithmical dependence on the parameter of
expansion. In this way, the theory dictates the form of $b_T$-behavior, although at the leading order the $b_T$-behavior is almost of our
choice.

OPE in the theory of interacting fields contains operators with more then two fields. In order to define the coefficient functions
self-consistently, the consideration of operators with the same dimension should be performed simultaneously. Since we are interested in the
operators of arbitrary high dimension, we should consider all possible operators. However, at a given order of the perturbative expansion the
number of operators is limited. In particular, for the definition of the coefficient functions of operators with two fields at order
$\mathcal{O}(g^2)$ one needs to consider only two-field operators. While, for the definition of the same operators at $\mathcal{O}(g^4)$ one
needs to consider additionally four-field operators (and, hence, four-point Green function of the non-local operator). The details of this
analysis are presented in the appendix A.

We limit our-self to the definition of the coefficient functions of the order $\mathcal{O}(g^2)$. As it is shown in the appendix A, we need to
consider only the two-point Green function of the operator $O(x,b_T)$. Since we are interested in the higher-derivative terms and our operator
breaks Lorentz invariance, the external fields of Green function should have an arbitrary momentum. Let us introduce the notation for the
two-point Green function and its perturbative expansion
\begin{eqnarray}\label{phi3:GreenF_notation}
Z_\phi(\mu)\langle \tilde \phi_r(p)O(x,b_T)\tilde \phi_r(p)\rangle&=&
\\\nn\mathcal{G}(x,b_T,p,\mu)&=&\mathcal{G}^{[0]}(x,b_T,p)+\alpha_g(\mu)
\mathcal{G}^{[2]}(x,b_T,p,\mu)+\mathcal{O}(\alpha_g^2),
\end{eqnarray}
where $\alpha_g(\mu)=g^2(\mu)/(4\pi)^3$. Correspondingly, the notation for the two-point Green function of the operators $O_n$ of OPE reads
\begin{eqnarray}\label{phi3:GreenFn_notation}
Z_\phi(\mu)\langle \tilde \phi_r(p)O_n(x)\tilde \phi_r(p)\rangle=\mathcal{G}^{[0]}_n(x,p)+\alpha_g(\mu)
\mathcal{G}^{[2]}_n(x,p,\mu)+\mathcal{O}(\alpha_g^2).
\end{eqnarray}
Finally, the perturbative expansion of the coefficient function reads
\begin{eqnarray}\label{phi3:CoefF_notation}
C_n(x,b_T,\mu)=C^{[0]}_n(x,b_T)+\alpha_g(\mu) C^{[2]}_n(x,b_T,\mu)+\mathcal{O}(\alpha_g^2),
\end{eqnarray}
where the leading term can be found from the free-theory expression (\ref{phi3:OP_Taylor_free_all})
\begin{eqnarray}
C_n^{[0]}(x)=\delta(1-x).
\end{eqnarray}
The coefficient $b_T^{\mu_1}...b_T^{\mu_n}/n!$ in the definition of OPE (\ref{phi3:OPE_simple}) is not a part of the coefficient function.
Therefore, the coefficient function (\ref{phi3:CoefF_notation}) is a dimensionless scalar function.

In the appendix A it shown that the coefficient $C^{[2]}$ can be obtain by solving the relation (\ref{app:G^2}), which in the notations
(\ref{phi3:GreenF_notation}-\ref{phi3:CoefF_notation}) reads
\begin{eqnarray}\label{phi3:G=C0G2+C2G0}
&&\mathcal{G}^{[2]}(x,b_T,p,\mu)= \sum_{n=0}^\infty \int_x^1\frac{dz}{z} \frac{b_T^{\mu_1}...b_T^{\mu_n}}{n!}\(
C_n^{[0]}\(\frac{x}{z}\)\mathcal{G}^{[2]}_{n,\mu_1...\mu_n}\(z,p,\mu\)+C_n^{[2]}\(\frac{x}{z},\mu\)\mathcal{G}^{[0]}_{n,\mu_1...\mu_n}\(z,p\)\).
\end{eqnarray}
The expression for the $G^{[2]}$ is given by a single diagram shown in fig.1, it reads
\begin{eqnarray}\label{phi3:G2_int}
G^{[2]}(x,b_T,p,\mu)=i\mu^{2\epsilon}g^2\int \frac{d^dk}{(2\pi)^d}\frac{\delta\(x -\frac{k_+}{p_+}\)e^{-i(kb)_T}}{[k^2+i0]^2[(k-p)^2+i0]},
\end{eqnarray}
where $d=6-2\epsilon$ the parameter of dimensional regularization.

The straightforward evaluation of the integral (\ref{phi3:G2_int}) gives us
\begin{eqnarray}\label{phi3:bessel+i0}
&&G^{[2]}(x,b_T,p,\mu)=2i^{-1-\frac{d}{2}}\frac{g^2}{(4\pi)^\frac{d}{2}} \mu^{2\epsilon}\bar x~e^{-ix(b p)_T }\(\frac{(b_T^2+i0)}{4(x\bar
xp^2+i0)}\)^\frac{\epsilon}{2}K_\epsilon\(\sqrt{(x\bar xp^2+i0)(-b_T^2-i0)}\),
\end{eqnarray}
where $\bar x=1-x$, and $K$ is the modified Bessel function of the second kind. We bring special attention to the signs of $i0$'s in arguments
of the function (\ref{phi3:bessel+i0}), since in the neighborhood of the interesting to us point $b_T^2=0$ Bessel functions tend to change its
kind. In our case $b_T^2>0$ one has
\begin{eqnarray}\label{phi3:G2_bessel}
G^{[2]}(x,b_T,p,\mu)=\bar x \alpha_g\mu^{2\epsilon}e^{\epsilon\gamma_E}~i\pi e^{i\pi\epsilon}\(\frac{b_T^2}{4p^2x\bar x}\)^{\frac{\epsilon}{2}}
H_\epsilon^{(1)}\(\sqrt{x\bar x p^2b_T^2}\)e^{-ix(bp)_T },
\end{eqnarray}
where $H^{(1)}$ is the Hankel function of the first order, and the $\overline{MS}$ scheme is applied.

\begin{figure}[t]
\begin{center}
\includegraphics[width=0.15\textwidth]{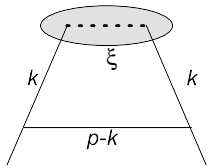}
\caption{One-loop graph needed for the evaluation of the coefficient function in $\phi^3_{D=6}$ theory.}
\end{center}
\end{figure}

The expression for $\mathcal{G}^{[2]}_n$ can be found by independent calculation of the loop-correction to the operator $O_n$ or, alternatively,
can be derived from the general expression (\ref{phi3:G2_bessel}). In the later case one needs to extract the term with the corresponding
(integer) power of $p$ and $b_T$ from (\ref{phi3:G2_bessel}) and then take the limit $b_T\to 0$.

The limit $b_T\to 0$ should be taken with special attention, because it does not commute with the limit $\epsilon\to 0$. One can reveal the
non-commutation of the limits by considering the series representation for the Hankel function. For the expression (\ref{phi3:G2_bessel}) the
expansion in the power of $p^2$ reads \cite{Bateman2}
\begin{eqnarray}\label{phi3:G2=series}
G^{[2]}(x,b_T,p,\mu)=\bar x \alpha_g e^{-ix(b p)_T}e^{\epsilon\gamma_E}\Bigg[\(\frac{\mu^2b^2_T}{4}\)^\epsilon\sum_{k=0}^\infty
\frac{\Gamma(-k-\epsilon)}{k!}\(\frac{x\bar x b_T^2p^2}{4}\)^k\\\nn+ e^{i\pi\epsilon}\(\frac{\mu^2}{x\bar x p^2}\)^\epsilon\sum_{k=0}^\infty
\frac{\Gamma(-k+\epsilon)}{k!}\(\frac{x\bar x b_T^2p^2}{4}\)^k\Bigg].
\end{eqnarray}
Indeed, taking limit $b_T\to0$, while keeping $\epsilon>0$, one eliminates all terms of the first sum in (\ref{phi3:G2=series}). The second sum
contains the poles of $\epsilon$ which should be subtracted in the operator of OPE. Taking the limits in the opposite order, i.e. $\epsilon\to
0$ and then $b_T\to 0$, one does not obtain the $\epsilon$-poles, but obtains the logarithms of $b_T$.

The difference between the order of limits results into the untrivial coefficient functions of OPE. Indeed, the solution of
(\ref{phi3:G=C0G2+C2G0}) can be schematically written as
\begin{eqnarray}\label{phi3:Ctrick}
C^{[2]}_n(x,b_T,\mu)=\(\lim_{b_T\to 0}\lim_{\epsilon\to 0}-\lim_{\epsilon\to 0}\lim_{b_T\to 0}\)\mathcal{G}^{[2]}\Big|_n,
\end{eqnarray}
where $\mathcal{G}\big|_n$ denotes the projection of the Green function onto the expression with the same Lorentz composition of $b_T$ and $p_T$
as in the operator $O_n$.

Considering the expression (\ref{phi3:G2=series}) one can see that our set of two-field operators (\ref{phi3:OP_Taylor_free_all}) is incomplete.
The set of two-field operators presented in (\ref{phi3:OP_Taylor_free_all}) contains only operators with derivatives in the transverse
directions. Whereas, the expression (\ref{phi3:G2=series}) contains the powers of $p^2$ as well. Therefore, the operators with different
combinations of $\partial^2$ operators should be added to the OPE. However, there are several observations which allows us to skip the
considerations of such operators.

First of all, the operators containing $(\partial^2)^n$ do not mix with operators which contains $(\partial^2)^k$ with $k<n$. This is a
consequence of the Lorentz invariance, e.g. an operator with $\partial^\mu_T\partial^\nu_T$ can mix with an operator $g_T^{\mu\nu}\partial^2$,
but not the opposite. Therefore, such operators do not influence on the definition of the coefficient function for operators with only
transverse derivatives.

Second, the matrix elements of operator with $\partial^2$ more suppressed in compare with matrix elements of only transverse derivatives.
Indeed, we can implement the counting rules $\partial_T\sim k_T$, while $\partial^2\sim \Lambda^2$. In the TMD regime $k_T^2\gg\Lambda^2$, the
matrix elements of the additional operators negligible.

Therefore, we omit operators with power of $\partial^2$ in OPE, although, in general, their contribution to the TMDs is non-zero. We note that
one should keep the non-zero $p^2$ during the calculation of the Green functions, in order to prevent artificial infrared singularities.

Applying the procedure (\ref{phi3:Ctrick}) we find the expressions for the coefficient functions
\begin{eqnarray}\label{phi3:coeff_Taylor}
C^{[2]}_n(x,b_T,\mu)=-\alpha_g x^n\bar x L_T,
\end{eqnarray}
where
\begin{eqnarray}\label{LT_def}
L_T=\ln\(\frac{\mu^2b_T^2}{4e^{-2\gamma_E}}\).
\end{eqnarray}
We note that the factor $e^{2\gamma_E}$ in the logarithm appear due to the finite part of the gamma-functions in the first sum
(\ref{phi3:G2=series}).

Thus, the expression for the Taylor-like OPE has the following form
\begin{eqnarray}\label{phi3:OP_Taylor_all}
&&O(x,b_T)=\\\nn&&\sum_{n=0}^\infty\int_x^1 \frac{dz}{z}\frac{b_T^{\mu_1}...b_T^{\mu_n}}{n!}\(\delta(1-z)-\alpha_g \bar x
x^nL_T+\mathcal{O}(\alpha_g^2)\) O_n^{\mu_1...\mu_n}\(\frac{x}{z},\mu\)+...~,
\end{eqnarray}
where
\begin{eqnarray}\label{phi3:OPE_op}
O_n^{\mu_1...\mu_n}\(x,\mu\)&=& p_+Z_\phi(\mu)\int\frac{d\xi^-}{2\pi} e^{-ixp_+ \xi^-}
\phi_r\(\xi^-/2\)\aboth{\partial^{\mu_1}_T}...\aboth{\partial^{\mu_n}_T}\phi_r\(-\xi^-/2\),
\end{eqnarray}
and the dots stay for the two-field operators with $\partial^2$ and  operators with more fields.

The $\mu$-dependence of the operator (\ref{phi3:OPE_op}) is given by the equation
\begin{eqnarray}\label{phi3:RG_Taylor}
\mu^2\frac{d}{d\mu^2}O_n^{\mu_1...\mu_n}\(x,\mu\)=\int_x^1 \frac{dz}{z}\[\alpha_g\(\bar z z^n-\frac{1}{6}\delta(1-z)\)
+\mathcal{O}(\alpha_g^2)\]O_n^{\mu_1...\mu_n}\(\frac{x}{z},\mu\),
\end{eqnarray}
where at $n=0$ the kernel is the DGLAP kernel in $\phi^3_{D=6}$, see e.g \cite{Mikhailov:1985cm}. The first term in the kernel comes from the
subtraction of the ultraviolet-singular term to the operator during the matching procedure (\ref{phi3:Ctrick}). It coincides (with the opposite
sign) with the coefficient infront of $\ln\mu^2$ in the coefficient function (\ref{phi3:coeff_Taylor}). So, in total, the $\mu$-dependence of
these terms cancels. The second term comes from the $\mu$-dependence of the field renormalization constant. And therefore, it resembles the
renormalization group kernel for the TMD operator.

Averaging over the directions and reorganizing the expansion with the help of (\ref{phi3:x^n->L_n(x)}) we obtain:
\begin{eqnarray}\label{phi3:OPE_Laguerre}
&&O(x,|b_T|)=\sum_{n=0,2,..}^\infty\int_x^1\frac{dz}{z}
\(\frac{z^2b_T^2}{4B_T^2}\)^\frac{n}{2}e^{-\frac{z^2b_T^2}{4B_T^2}}\(\delta(1-z)-\alpha_g\bar z L_T+\mathcal{O}(\alpha_g^2)\)
\mathbb{O}_n\(\frac{x}{z},B_T,\mu\)+...,
\end{eqnarray}
where
\begin{eqnarray*}
&&\mathbb{O}_n(x,B_T,\mu)=p_+\frac{\(\frac{d_\perp-2}{2}\)!}{\(\frac{n+d_\perp-2}{2}\)!}Z_\phi(\mu)\int\frac{d\xi^-}{2\pi} e^{-ixp_+ \xi^-}
\phi_r\(\xi^-/2\)L^\frac{d_\perp-2}{2}_\frac{n}{2}\(-\aboth{\partial^{2}_T}B_T^2\)\phi_r\(-\xi^-/2\),
\end{eqnarray*}
and the dots stay for the two-field operators with $\partial^2$ and  operators with more fields.

The $\mu$-dependence of the operators $\mathbb{O}_n$ can be obtained via $\mu$-dependence of the operators $O_n$ and the algebraic relation
between them. The operators $\mathbb{O}_n$ satisfy the following equation
\begin{eqnarray}\label{phi3:RG_Laguerre}
\mu^2\frac{d}{d\mu^2}\mathbb{O}_n\(x,\mu\)&=&\sum_{k=0,2,..}^n\int_x^1 \frac{dz}{z} P_{n,k}\(z,\alpha_g\)\mathbb{O}_k\(\frac{x}{z},\mu\),
\end{eqnarray}
where
\begin{eqnarray}
P_{n,k}(x,\alpha_g)=\alpha_g\(\frac{x^k(1-x^2)^\frac{n-k}{2}}{\(\frac{n-k}{2}\)!}-\frac{\delta_{n,k}}{6}\delta(1-x)\)+\mathcal{O}\(\alpha_g^2\).
\end{eqnarray}
At $n=0$ the equation (\ref{phi3:RG_Laguerre}) is the DGLAP equation.

Taking the matrix element of the operator in the hadronic brackets we obtain the modified small-$b_T$ expansion for TMD PDF
\begin{eqnarray}\label{phi3:TMD_PDF}
F(x,b_T,\mu)=\int_x^1\frac{dz}{z}\(\delta(1-z)e^{-\frac{b_T^2}{4B_T^2}}-\alpha_g \bar x
L_Te^{-\frac{z^2b_T^2}{4B_T^2}}+\mathcal{O}(\alpha^2_g)\)f\(\frac{x}{z},\mu\)+...,
\end{eqnarray}
where $f(x,\mu)$ is the integrated PDF and the dots denote terms involving the higher order operators.

The expression (\ref{phi3:TMD_PDF}) is singular at $b_T=0$. The singularity appears because in the regime $b_T\to0$ the TMD factorization should
be replace by the collinear factorization. Different factorizations sums up different parts of the original OPE of the hadronic tensor.
Therefore, the one should perform additional matching procedure in order to join the regimes of collinear factorization (small-$b_T$) and TMD
factorization (middle-$b_T$).

The matching procedure can be done by performing OPE on the scale different from $\mu$, e.g. on the scale $\kappa$ like in the expression
(\ref{phi3:OPE_simple}). The scale $\kappa$ should be kept in the range from $\Lambda^2$ and $b_T^{-2}$ in order to prevent the logarithms of
$(b_T\kappa)$ be large. Since, in this regime the logarithms of $(\kappa/\mu)$ in the coefficient functions became large and they should
resummed with the help of DGLAP equation.

Another way of matching the regimes is to keep the logarithms $L_T$ small at small-$b_T$ by setting the renormalization scale $\mu\sim b_T^{-1}$
at $b_T\to 0$. The most popular choice of the scale $\mu_b$ was suggested in \cite{Collins:1981va} and reads
\begin{eqnarray}\label{mu_b}
\mu^2_b=4e^{-2\gamma_E}\(\frac{1}{b_T^2}+\frac{1}{b_{\text{max}}^2}\),
\end{eqnarray}
where $b_{\text{max}}$ should be chosen such that $\alpha(\mu_b)_{b_T\to \infty}$ and $L_T$ have reasonably small values. Then performing
evolution of the TMD PDF down to the scale $\mu_b$ we obtain
\begin{eqnarray}\label{phi3:TMD_PDF_mub}
F(x,b_T,\mu)&=&\exp\[\int_{\mu^2_b}^{\mu^2}\frac{d\mu'^2}{\mu'^2}\(\frac{\alpha_g(\mu')}{6}+\mathcal{O}(\alpha_g^2)\)\]
\\\nn&&\times\int_x^1\frac{dz}{z}\(\delta(1-z)e^{-\frac{b_T^2}{4B_T^2}}-\alpha_g(\mu_b) \bar z
\ln\(\frac{b_T^2\mu^2_b}{4e^{-2\gamma_E}}\)e^{-\frac{z^2b_T^2}{4B_T^2}}+\mathcal{O}(\alpha^2_g)\)f\(\frac{x}{z},\mu_b\).
\end{eqnarray}
This expression is finite at $b_T\to 0$ for finite $x$.

The correction to the leading result is larger at smaller $x$. One can see that the dominating behavior in the form of the Gaussian factor with
$x^2$ in the argument, would appear in all orders of the perturbation expansion. This indicates that there is a natural bound of convergence for
Laguerre-based OPE. The bound is dependent on $x$ and $B_T$ and follows from the demand that the correction should be smaller in comparison with
the leading term. In the case of $\phi^3_{D=6}$ this requirement leads to $\alpha_g(\mu_b)L_T<\exp\((x^2-1)b_T^2/4B_T^2\)$.

\section{Laguerre-based expansion for unpolarized TMD PDF}
\subsection{TMD operator definition}
The gauge invariance of QCD leads to significant complication of the TMD factorization procedure. The soft gluon exchange between hadron states
results to additional logarithms to be factorized. On the operator level, the gauge invariance is achieved by Wilson lines connecting the quark
fields. The geometry of the Wilson lines is dependent on the details of process kinematics.

The factorization of the parton distributions related to different hadron states leads to additional divergences, called rapidity divergences,
in every TMD operator. The cancelation of the rapidity divergences takes a place inside the product of TMDs and the soft factor. The soft factor
is given by the vacuum expectation of the Wilson lines combination and also depends on the kinematics of the TMD process, see e.g.
\cite{Collins:2011zzd,Ji:2004wu}. In order to have the divergence-free definition of TMDs, the soft factor can be included into the definition
of TMDs operator in the form of the renormalization multiplier \cite{Cherednikov:2007tw,Collins:2011ca,Echevarria:2012js}.

The definition of the TMD PDF is dependent on the kinematics of the process. In the following we consider the Drell-Yan kinematics and TMD PDF
for the hadron propagating along the direction $n$. Then, the quark TMD PDF reads
\begin{eqnarray}
f_{q/H}(x,b_T;\mu,\zeta_n)&=&\langle H(p)|O_q(x,b_T;\mu,\zeta_n)|H(p)\rangle,
\end{eqnarray}
where $\mu$ is the renormalization and hard factorization scale, and $\zeta_n$ is the scale of rapidity divergences subtraction. The quark TMD
operator reads
\begin{eqnarray}\label{QCD:Op}
&&O_q(x,b_T;\mu,\zeta_n)=Z_q(\mu)S^{-\frac{1}{2}}\!(b_T,\zeta_n) \int\frac{d\xi^-}{2\pi}e^{-ix p^+\xi^-}\bar
q_r\!\!\(\frac{\xi}{2}\)W^\dagger\!\!\(\frac{\xi}{2},-\infty; n\)\frac{\gamma^+}{2}W\!\!\(-\frac{\xi}{2},-\infty;
n\)q_r\!\!\(-\frac{\xi}{2}\)\!\!,
\end{eqnarray}
where $\xi=\{0,\xi^-,b_T\}$, and $W$ denotes the Wilson line
$$
W\(x,-\infty; n\)=P\exp\(-ig\int_{-\infty}^0 d\sigma ~n^\mu A_\mu(x+\sigma n)\).
$$
The factor $Z_q$ is the field renormalization constant, while the factor $S$ is the rapidity divergences subtraction factor.

The factor $S$ in (\ref{QCD:Op}) can be defined in different ways. Two major definitions are so-called, Collins' definition
\cite{Collins:2011ca,Collins:2011zzd}, and so-called, Echevarria-Idilbi-Scimemi definition \cite{GarciaEchevarria:2011rb}. These approaches are
equivalent \cite{Echevarria:2012js,Collins:2012uy}, although they are quite different in the form. In the following we are going to use the
Echevarria-Idilbi-Scimemi definition, in which the factor $S$ is defined as \cite{Echevarria:2012js}
\begin{eqnarray}\label{QCD:deltapm}
S(b_T,\zeta_n)=\tilde S\(\frac{\Delta}{p^-},\alpha\frac{\Delta}{p^+}\),
\end{eqnarray}
where $\alpha=2 p^+p^- \zeta_n$ and $\Delta$ is the parameter of the $\delta$-regularization \cite{Chiu:2009yx,GarciaEchevarria:2011rb}, see
also appendix B. The $\delta$-regularization is used for the regularization of the rapidity divergences and the dependence on the parameter
$\Delta$ is canceled in the product (\ref{QCD:Op}). The expression for the $\tilde S$ is given by the following vacuum matrix element
\begin{eqnarray}
&&\tilde S(\delta^+,\delta^-)=\langle0|\frac{1}{N_c}\tr\(W^\dagger\!\!\(\frac{b_T}{2},-\infty;n\)W\!\!\(\frac{b_T}{2},-\infty;\bar n\)
W^\dagger\!\!\(-\frac{b_T}{2},-\infty;\bar n \)W\!\!\(-\frac{b_T}{2},-\infty;n\)\)\!\!|0\rangle.
\end{eqnarray}
The parameters $\delta^\pm$ regularize the rapidity divergences of $W(n)$ and $W(\bar n)$, respectively.

\subsection{Coefficient functions in the Laguerre basis}

In order to obtain OPE in the Laguerre polynomial basis, we use the same strategy as for the $\phi^3_{D=6}$ model. First, we obtain the
Taylor-like expression for the OPE. Second, we reorganize it in the Laguerre polynomials.

In order to obtain the coefficient functions for the operators with two fields at the order $\mathcal{O}(g^2)$ we need to calculate the
two-point Green function
\begin{eqnarray}\label{QCD:G2_tree}
Z_q(\mu)\langle q_r(p) O_q(x,b_T;\mu,\zeta_n) \bar q_r(p) \rangle=\mathcal{G}_{q/q}^{[0]}(x,b_T,p)+a_s
\mathcal{G}_{q/q}^{[2]}(x,b_T,p)+\mathcal{O}(a_s^2),
\end{eqnarray}
where $a_s=\frac{g^2}{(4\pi)^2}=\frac{\alpha_s}{4\pi}$. The calculation of the function $\mathcal{G}^{[2]}_{q/q}$ is presented in the appendix
B.

In QCD case OPE contains operators with various Lorentz structures. There are operators with $\gamma^-$ and $\gamma_T^\mu$, and various
components of derivatives. Since we consider only unpolarized TMDs, we can neglect many of such operators, because they vanish after averaging
over directions. The final expression contains three types of operators. They are the operators with $\gamma^+(\partial_T^2)^n$,
$\not{\!\partial}_T(\partial_T^2)^n$, and $\gamma^-(\partial_T^2)^n$ (there are also operators with $\partial^2$, but they can be omitted due to
the same reasons as in $\phi^3_{D=6}$, see discussion after (\ref{phi3:Ctrick})). The operators belonging to these three types do not mix with
each other due to different Lorentz transformation properties. Also the matrix elements of operators with $\gamma_T$ and $\gamma^-$ are
suppressed due to the equation of motions (by $\Lambda/p^+$ and $(\Lambda/p^+)^2$ respectively). Therefore, we omit these operators at our level
of accuracy.

There are also pure gluon operators which contribute to the OPE. The coefficient function of gluon operators can be found by evaluation of the
Green function with external gluon fields
\begin{eqnarray}\label{QCD:G2_tree}
Z_g(\mu)\langle A^\mu_r(p) O_q(x,b_T;\mu,\zeta_n) A^\nu_r(p) \rangle=a_s \mathcal{G}^{\mu\nu[2]}_{q/G}(x,b_T,p)+\mathcal{O}(a_s^2).
\end{eqnarray}
The amount of various Lorentz structures for the gluon operator is larger. However, the most of operators do not contribute to the unpolarized
TMD, and do not mix with the leading operators. Therefore, at our level of accuracy we can average over the polarizations of gluons. The
calculation of the function $\mathcal{G}_{g/q}$ is presented in appendix B.

Thus, the OPE for the TMD operator (\ref{QCD:Op}) reads
\begin{eqnarray}\label{QCD:OPE_Taylor}
O_q(x,b_T;\mu,\zeta_n)&=&\sum_{n=0}^\infty\int_x^1\frac{dz}{z} C_{q/q}^{\mu_1...\mu_n}(z,b_T,\mu,\zeta)O_q^{\mu_1...\mu_n}\(\frac{x}{z},\mu\)
\\\nn&&+\sum_{n=0}^\infty\int_x^1\frac{dz}{z} C_{q/g}^{\mu_1...\mu_n}(z,b_T,\mu,\zeta)O_g^{\mu_1...\mu_n}\(\frac{x}{z},\mu\)+...~,
\end{eqnarray}
where the operators are
\begin{eqnarray}\nn
 O_q^{\mu_1...\mu_n}\(x,\mu\)&=&Z_q(\mu)\int\frac{d\xi^-}{2\pi}e^{-ixp_+\xi^-}\\\nn&&\times\bar q_r\(\frac{\xi^-}{2}\)
W^\dagger\(\frac{\xi^-}{2},-\infty; n\)\frac{\gamma^+}{2}\aboth{\partial_T^{\mu_1}}...\aboth{\partial_T^{\mu_n}}W\(-\frac{\xi^-}{2},-\infty;
n\)q_r\(\frac{-\xi^-}{2}\),
\\\nn
O_g^{\mu_1...\mu_n}\(x,\mu\)&=&Z_g(\mu)\int\frac{d\xi^-}{2\pi}e^{-ixp_+\xi^-}\\\nn&& \times G^{+\alpha}_r\(\frac{\xi^-}{2}\)
W^\dagger\(\frac{\xi^-}{2},-\infty; n\)\aboth{\partial_T^{\mu_1}}...\aboth{\partial_T^{\mu_n}}W\(-\frac{\xi^-}{2},-\infty;
n\)G_r^{+\alpha}\(\frac{-\xi^-}{2}\).
\end{eqnarray}
The dots in (\ref{QCD:OPE_Taylor}) denote the operators with more then two fields (without counting the Wilson lines), and the operators with
other Lorentz structures. The coefficient functions for the individual terms of OPE can be obtained with the help of (\ref{phi3:Ctrick}).

These operators are gauge-invariant in any non-singular gauge, because the parts of operators on the left- and the right-sides of the transverse
derivatives are gauge-invariant independently. In the singular gauges, such as the light-cone gauge, transverse derivatives $\partial^\mu_T$
should be extended to a covariant derivative at light-cone infinity: $\partial^\mu_T-i g A^\mu_T(-n\infty)$. These additional transverse gluons
are produced by the expansion of the transverse links at light-cone infinity, which make the definition of the TMD PDF gauge-invariant
\cite{Belitsky:2002sm}.

The coefficient functions of (\ref{QCD:OPE_Taylor}) can be calculated in the same way as in the $\phi^3_{D=6}$ theory. The intermidiate
expressions for the Green function are presented in appendix B.
\begin{eqnarray}\label{QCD:Cqq_Taylor}
C_{q/q}^{\mu_1...\mu_n}(x,b_T,\mu,\zeta)&=& b_T^{\mu_1}...b_T^{\mu_n}\frac{\delta(1-x)}{n!}
\\\nn&&+2a_sC_F\frac{x^n}{n!}\Bigg\{
b_T^{\mu_1}...b_T^{\mu_n}\bigg[
 -L_T P_{qq}(x)+\bar x-n\bar x\(L_T-1\)
 \\\nn&&\qquad\qquad
 +\delta(\bar x)\(\frac{3}{2}L_T-\frac{1}{2}L_T^2-\frac{\pi^2}{12}+L_T\ln\(\frac{\mu^2}{\zeta}\)\)\bigg]
\\\nn&&\qquad\qquad+
b_T^{\mu_1}...b_T^{\mu_{n-2}}b_T^2g^{\mu_{n-1}\mu_{n}}n(n-1)\(2-L_T\)\Bigg\} +\mathcal{O}(a_s^2),
\end{eqnarray}
\begin{eqnarray}\label{QCD:Cqg_Taylor}
C_{q/q}^{\mu_1...\mu_n}(x,b_T,\mu,\zeta)&=&2a_s\frac{x^n}{n!}b_T^{\mu_1}...b_T^{\mu_n}\(-P_{qg}(x)L_T+2x\bar x\)+\mathcal{O}(a_s^2),
\end{eqnarray}
where $L_T$ is defined in (\ref{LT_def}) and the functions $P$ are the kernels of DGLAP equation
$$
P_{qq}(x)=\(\frac{1+x^2}{1-x}\)_+,~~~~~~P_{qg}(x)=1-2x\bar x.
$$
At $n=0$ coefficient function (\ref{QCD:Cqq_Taylor},\ref{QCD:Cqg_Taylor}) were calculated in many papers, e.g. see
\cite{Collins:2011zzd,Aybat:2011zv,Bacchetta:2013pqa,GarciaEchevarria:2011rb}. In the kinematics of semi-inclusive deep inelastic scattering the
term $\pi^2/12$ in the second line of (\ref{QCD:Cqq_Taylor}) disappears.

The $\mu$-dependence of the transversally local operators is given by the DGLAP equation
\begin{eqnarray}\label{QCD:DGLAP}
\mu^2\frac{d
O_{i}^{\mu_1...\mu_n}(x,\mu)}{d\mu^2}=\int_x^1\frac{dz}{z}\(P^n_{ij}(x)\)^{\mu_1...\mu_n}_{\nu_1...\nu_n}O_{j}^{\nu_1...\nu_n}\(\frac{x}{z},\mu\),
\end{eqnarray}
where the kernels are
\begin{eqnarray}
P^n_{qq}(x)&=&a_sx^n\delta^{\mu_1}_{\nu_1}...\delta^{\mu_{n-2}}_{\nu_{n-2}}\[\(P_{qq}(x)+n\bar
x\)\delta^{\mu_{n-1}}_{\nu_{n-1}}\delta^{\mu_n}_{\nu_n}-\frac{n(n-1)}{2}g^{\mu_{n-1}\mu_n}g_{\nu_{n-1}\nu_n}\],
\\
P^n_{qg}(x)&=&a_sx^n\delta^{\mu_1}_{\nu_1}...\delta^{\mu_{n}}_{\nu_{n}}P_{qg}(x).
\end{eqnarray}

Averaging over the directions of the $b_T$, and applying the operator redefinition (\ref{phi3:x^n->L_n(x)}) we obtain the OPE in the Laguerre
basis
\begin{eqnarray}\label{QCD:OPE_Laguerre}
O_q(x,b_T;\mu,\zeta_n)&=&\sum_{n=0,2,..}^\infty\int_x^1\frac{dz}{z} \mathbb{C}^n_{q/q}(z,b_T,\mu,\zeta)\mathbb{O}^n_q\(\frac{x}{z},\mu\)
\\\nn&&+\sum_{n=0,2..}^\infty\int_x^1\frac{dz}{z} \mathbb{C}^n_{q/g}(z,b_T,\mu,\zeta)\mathbb{O}^n_g\(\frac{x}{z},\mu\)+...~,
\end{eqnarray}
where the coefficient functions are
\begin{eqnarray}\label{QCD:Laguerre_Cqq}
 \mathbb{C}^n_{q/q}(x,b_T,\mu,\zeta)&=&\(\frac{b_T^2}{4B_T^2}\)^\frac{n}{2}e^{-\frac{b_T^2}{4B_T^2}}\delta(1-x)
 \\ \nn &&+
 2a_s C_F\(\frac{x^2b_T^2}{4B_T^2}\)^\frac{n}{2}e^{-\frac{x^2b_T^2}{4B_T^2}}\Bigg[
 \\\nn &&\qquad-L_TP_{qq}(x)+\delta(\bar
x)\(\frac{3}{2}L_T-\frac{1}{2}L_T^2-\frac{\pi^2}{12}+L_T\ln\(\frac{\mu^2}{\zeta}\)\)
\\\nn &&\qquad\quad-\frac{\bar x}{4} L_T\(\frac{x^4}{4}\(\frac{b_T^2}{B_T^2}\)^2-(n+3)x^2\frac{b_T^2}{B_T^2}+n(n+4)\)
\\\nn &&\qquad\quad+\frac{\bar x}{4}\(\frac{x^4}{2}\(\frac{b_T^2}{B_T^2}\)^2-2(n+2)x^2\frac{b_T^2}{B_T^2}+2n(n+2)\)
\Bigg]+\mathcal{O}(a_s^2),
\\\label{QCD:Laguerre_Cqg}
 \mathbb{C}^n_{q/g}(x,b_T,\mu,\zeta)&=& 2a_s \(\frac{x^2b_T^2}{4B_T^2}\)^\frac{n}{2}e^{-\frac{x^2b_T^2}{4B_T^2}}\(-P_{qg}(x)L_T+2x\bar
 x\)+\mathcal{O}(a_s^2).
\end{eqnarray}
The evolution equations for the operators $\mathbb{O}^n$ can be easily derived, and they have the DGLAP structure.

\subsection{Modified expression for TMD PDF}
With the help of the OPE (\ref{QCD:OPE_Laguerre}) we can obtain the modified expression for the TMD PDF, which is presumably valid in some wide
region of $b_T$. It reads
\begin{eqnarray}\label{QCD:F=f}
F_{q/H}(x,b_T;\mu,\zeta)&=&\sum_{j}\int_x^1\frac{dz}{z}\mathbb{C}_{q/j}\(\frac{x}{z},b_T;\mu,\zeta\)f_{j/H}(z,\mu)+\mathcal{O}_1,
\end{eqnarray}
where the symbol $\mathcal{O}_1$ denotes the order of eliminated contribution. The estimation of the size for $\mathcal{O}_1$ is impossible
within the perturbative QCD, but can be obtained from the comparison with experimental data or other theoretical approaches. In the following we
suppose that $\mathcal{O}_1$ is negligible in comparison with the first term of (\ref{QCD:F=f}).

The function $f_{j/H}$ in (\ref{QCD:F=f}) is integrated PDF. The coefficient function $\mathbb{C}$ can be obtained from equations
(\ref{QCD:Laguerre_Cqq}-\ref{QCD:Laguerre_Cqg}) at $n=0$,
\begin{eqnarray}\label{QCD:Laguerre_Cqq_n=0}
 \mathbb{C}_{q/q}(x,b_T,\mu,\zeta)&=&e^{-\frac{b_T^2}{4B_T^2}}\delta(1-x)
 \\ \nn &&+
 2a_s C_Fe^{-\frac{x^2b_T^2}{4B_T^2}}\Bigg[-L_TP_{qq}(x)+\bar x
 \\\nn&&\qquad+\delta(\bar
x)\(\frac{3}{2}L_T-\frac{1}{2}L_T^2-\frac{\pi^2}{12}+L_T\ln\(\frac{\mu^2}{\zeta}\)\)
\\\nn&&\qquad-\frac{\bar x x^2}{4} \frac{b_T^2}{B_T^2}L_T\(\frac{x^2}{4}\frac{b_T^2}{B_T^2}-3\)+\frac{x^4\bar x}{8}\(\frac{b_T^2}{B_T^2}\)^2
-x^2\bar x\frac{b_T^2}{B_T^2} \Bigg]+\mathcal{O}(a_s^2),
\\\label{QCD:Laguerre_Cqg_n=0}
 \mathbb{C}_{q/g}(x,b_T,\mu,\zeta)&=& 2a_s e^{-\frac{x^2b_T^2}{4B_T^2}}\(-P_{qg}(x)L_T+2x\bar
 x\)+\mathcal{O}(a_s^2).
\end{eqnarray}
At $b_T\to 0$ this expressions reveal the standard expressions for the matching coefficients of TMD PDF to integrated PDF. For simplicity we
take the size of the Gaussian distribution $B_T$ the same for the quark and gluon contributions. However, due to algebraic independence of these
parts of OPE, the parameters $B_T$ can be different for these distributions.

The derived expression can be compared to the standard expression used in the phenomenology (see e.g.
\cite{Aidala:2014hva,Aybat:2011zv,Aybat:2011ge,Bacchetta:2013pqa})
\begin{eqnarray}\label{QCD:F=standard}
&&F_{q/H}(x,b_T;\mu,\zeta)=\sum_{j}\int_x^1\frac{dz}{z}C_{q/j}\(\frac{x}{z},b_T;\mu,\zeta\)f_{j/H}(z,\mu)\exp\(g_{j/H}(x,b_T)+g_K(b_T)\ln\zeta\),
\end{eqnarray}
where $g$'s are unknown functions, and $C_{q/j}$ is the coefficient function for Taylor-like expansion at $n=0$
(\ref{QCD:Cqq_Taylor}-\ref{QCD:Cqg_Taylor}). The typical expression for the function $g$ is $g=-b_T^2\langle P_T^2\rangle/4$. One can see that
expansion over the Laguerre basis at $\mathcal{O}(a_s^2)$ implies the following expression for functions $g$
\begin{eqnarray}\label{QCD:g}
g_{q/H}\sim-x^2\frac{b_T^2}{4B_T^2}\ln\[1-a_s\frac{\bar x x^2}{4}\frac{b_T^2}{B_T^2}\(
L_T\(\frac{x^2}{4}\frac{b_T^2}{B_T^2}-3\)-\frac{x^2}{2}\frac{b_T^2}{B_T^2} +4\)\],
\end{eqnarray}
which should be considered as a part of coefficient function (i.e. in the convolution with integrated PDF). The main difference of this
expression from the frequently used Gaussian is the common multiplier $x^2$. This multiplier spreads the coefficient function at smaller $x$.
The logarithm factor in (\ref{QCD:g}) produces small relative effect in the region $b_T<B_T$, where our calculation is applicable.

The TMD PDF is dependent on the scale parameters $\mu$ and $\zeta$. This dependence is given by the Collins-Soper equation and renormalization
group equation. Together they result to the equations
\begin{eqnarray}\label{QCD:RG}
\mu^2\frac{d}{d\mu^2}F(x,b_T;\mu,\zeta)&=&\(2a_sC_F\(\frac{3}{2}-\ln\(\frac{\zeta}{\mu^2}\)\)+\mathcal{O}\(a_s^2\)\)F(x,b_T;\mu,\zeta),\\\label{QCD:CS}
\zeta\frac{d}{d\zeta}F(x,b_T;\mu,\zeta)&=&\(-2a_sC_FL_T+\mathcal{O}\(a_s^2\)\)F(x,b_T;\mu,\zeta).
\end{eqnarray}
On the right-hand side of the expression (\ref{QCD:Laguerre_Cqq_n=0}) such behavior is not transparent. The $\mu$-dependence of the coefficient
function is collected from the general $\mu$-dependence of the TMD operator (\ref{QCD:Op}) and the $\mu$-dependence induced by OPE. The later
cancels between the operators and the coefficient functions in the expression (\ref{QCD:OPE_Laguerre}), so the resulting $\mu$-dependence of the
operator (\ref{QCD:Op}) and its OPE coincide.

In the Taylor-like OPE every subset of operators with the same canonical dimension satisfies the same renormalization group equations as the
operator. This is not the case of the Laguerre-based OPE, because the later are constructed from the operators of different dimensions.
Therefore, the $\mu$-dependence and $\zeta$-dependence of every individual term of Laguerre-based OPE should not reproduce these dependencies of
TMD operator, as it takes a place in Taylor-like OPE. Nonetheless, one can see from (\ref{QCD:OPE_Laguerre}), that the $\zeta$-dependence is
universal for all terms of OPE, as well as $b_T$-independent part of $\mu$-dependence. These terms of OPE (they are collected in the first and
second line of (\ref{QCD:Laguerre_Cqq_n=0}), and the expression (\ref{QCD:Laguerre_Cqg_n=0}) in total)  behave in the usual way, i.e. every term
reproduces the renormalization group behavior of the initial operator. The $\mu$-dependence of the rest terms (the third line of
(\ref{QCD:Laguerre_Cqq_n=0})) cancels only in the sum of all operators. Therefore, during the evolution with respect to $\mu$ some portion of
the truncated expression (\ref{QCD:F=f}) ``flows away'' to the higher order terms. However, this ``lost'' part is of order $\mathcal{O}_1$ so it
can be neglected in the range of application of expression (\ref{QCD:F=f}).

The OPE (\ref{QCD:OPE_Laguerre}) is performed at such a scale that keeps the logarithms $L_T$ small. The common choice of the factorization
scale $\mu$ is $\mu_b$, introduced in (\ref{mu_b}). Then the value of $L_T$ is weakly dependent on $b_T$, and small in the region $b_T \lesssim
b_{\text{max}}$. The value of $b_\text{max}$ is typically taken of order $1$ GeV$^{-1}$, and characterizes the boundary of the perturbative
region. Simultaneously, the parameter $\zeta$ should be taken such that $\ln\(\zeta/\mu_b^2\)$ is also small. The best choice is $\zeta=\mu_b$.

In practice, the values of parameters $\mu$ and $\zeta$ are dictated by the TMD factorization theorem, which results to $\mu^2\sim \zeta\sim
Q^2$. Therefore, in order to apply the OPE we first need to evolve the TMD PDF from $\mu^2\sim \zeta\sim Q^2$ down to $\mu^2=\zeta=\mu_b^2$. The
evolution is given by equations (\ref{QCD:RG},\ref{QCD:CS}). We obtain
\begin{eqnarray}\label{QCD:F=f_evol}
F_{q/H}(x,b_T;\mu,\zeta)&=&\sum_{j}\int_x^1\frac{dz}{z}\mathbb{C}_{q/j}\(\frac{x}{z},b_T;\mu_b,\mu_b^2\)f_{j/H}(x,\mu_b)
\\ &&\nn\times
\exp\[-2a_s(\mu_b)C_F\ln\(\frac{\zeta}{\mu_b^2}\)L^b_T+2C_F\int_{\mu_b}^\mu
\frac{d\mu'^2}{\mu'^2}a_s(\mu')\(\frac{3}{2}-\ln\(\frac{\zeta}{\mu'^2}\)\)\],
\end{eqnarray}
where $L_T^b=\ln\(b_T^2\mu_b^2/4e^{-2\gamma_E}\)$.

The TMD evolution is independent on $x$. Indeed, since the kernels of equations (\ref{QCD:RG}-\ref{QCD:CS}) are independent on $x$, the
evolution exponent in (\ref{QCD:F=f_evol}) is also $x$-independent. Therefore, we have two free parameters, namely, $B_T$ and $b_\text{max}$.

\begin{figure}[t]
\includegraphics[width=0.25\textwidth]{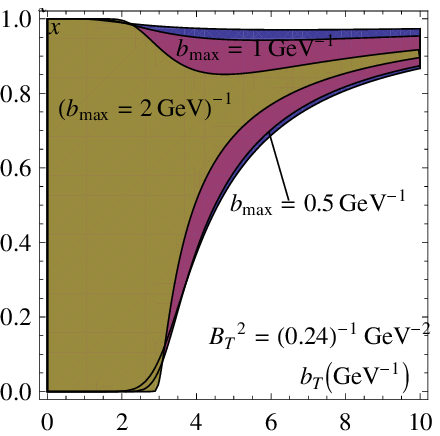}~~~\includegraphics[width=0.25\textwidth]{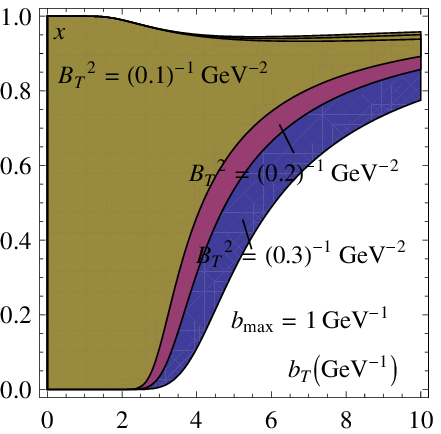}~~~\includegraphics[width=0.4\textwidth]{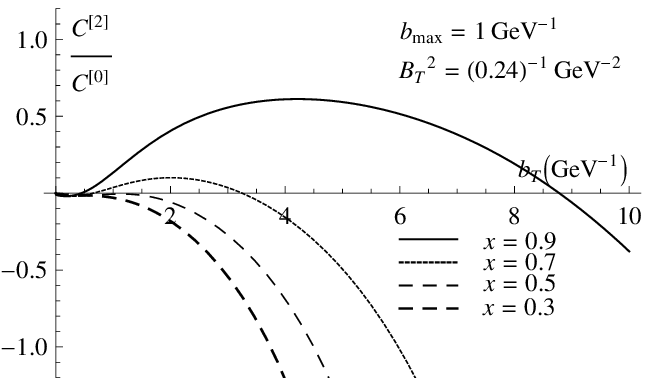}
\caption{Plot of the ratios of the coefficient functions $C^{[2]}_{q/q}$  and $C^{[0]}_{q/q}$(in convolution with a constant test function).
(Left and central panels) The regions of $C^{[2]}<C^{[0]}$ at different $b_\text{max}$(left) and $B_T$(center). (Right panel) The profile of the
ratio $C^{[2]}/C^{[0]}$ at different $x$.} \label{fig:ratio}
\end{figure}

The value of the $b_\text{max}$ can be fixed from two assumptions. First assumption is that $a_s(\mu_b)$ should not touch the Landau singularity
at any $b_T$ (this assumption restricts $b_\text{max}$ from above). Second assumption is that $L_T^b$ should be reasonably small in some wide
region of $b_T$ (this assumption restricts $b_\text{max}$ from below). These assumptions lead to the typical range of $b_\text{max}$ from $0.5$
GeV$^{-1}$ to $2$ GeV$^{-1}$.

We have an additional restriction on $b_\text{max}$. The calculated correction should be reasonably small in the comparison with the leading
term. Since the coefficient function is the generalized function for the comparison we convolute it with some test function. In
fig.\ref{fig:ratio}(left panel) one can see the regions where the correction to the coefficient function is less then the leading order. The
region of small correction is wider at large-$x$, and it is not very sensitive to the variation of $b_\text{max}$. Therefore, in the following
we choose $b_\text{max}=1$ GeV$^{-1}$.

In our approach the parameter $B_T$ cannot be fixed from any principle. The recent phenomenological investigations of TMDs at low $Q$
\cite{Aidala:2014hva} suggest the width of the Guassian exponent $B_T^2\simeq (0.24)^{-1}$ GeV$^{-2}$ (for $Q^2\simeq 2.1$ GeV$^2$). For such
values the relative size of correction is reasonably small up to $b_T\sim2$ GeV$^{-1}$ for small $x$ and up to $b_T\sim 6-7$ GeV$^{-1}$ for
large $x$. It follows from the consideration of the ratio of coefficient functions, which is shown in fig.\ref{fig:ratio}(right panel).

The general form of the validity region can be understood from the central panel in fig.\ref{fig:ratio}, where we show the region of small
perturbative correction for different values of $B_T$. For these estimations we have used the constant as a test function. Usage of other
test-functions  results to the similar, but less restrictive conclusions.

\begin{figure}[t]
\includegraphics[width=0.3\textwidth]{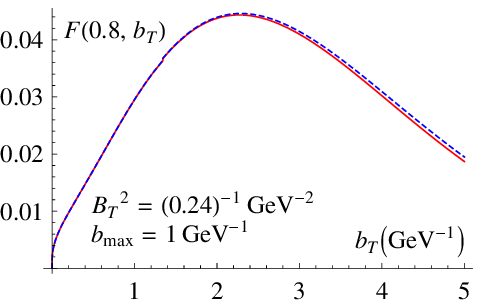}~~~\includegraphics[width=0.3\textwidth]{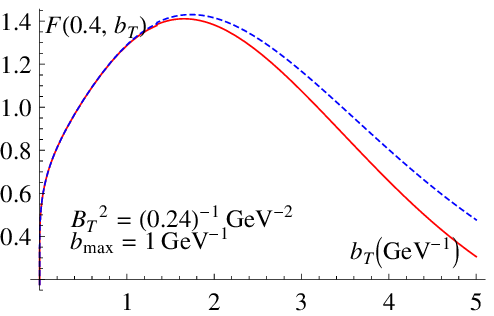}~~~\includegraphics[width=0.3\textwidth]{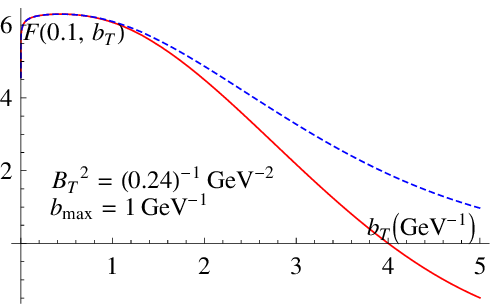}
\caption{Plots of TMD PDF given in (\ref{QCD:F=f})(red curve) at different values of $x$ ($x=0.8,0.4,0.1$ from left to right panels). The
blue-dashed curves are the TMD PDF calculated in Taylor-like basis multiplied by a non-perturbative factor $\exp(-b_T^2/4B_T^2)$. The evolution
exponent in omitted.} \label{fig:TMDs}
\end{figure}

The character feature of the result is that the obtained correction is negative. This is the general result, which can be deduced without
calculation. Indeed, the main contribution to the coefficient function comes from the term proportional to $L_T$, which, in its own turn, is
proportional to DGLAP kernel with the negative sign (we remind that the negative sign is the result of the general renormalization point
independence of OPE). Since the DGLAP kernel is strictly positive, the main correction to coefficient function is negative. At some very large
$b_T$ the power corrections became significant and they can result into the positive expression. In fig.\ref{fig:TMDs} we compare our expression
with the expression (\ref{QCD:F=standard}) at $b_\text{max}=1$GeV$^{-1}$, $B_T^2=(0.24)^{-1}$GeV$^{-2}$, and $g=-b_T^2/4B_T^2$. The deviation
from the standard approach is small for the large-$x$, even for very large $b_T$. While, for the smaller-$x$ the difference between the
expressions (\ref{QCD:F=f}) and (\ref{QCD:F=standard}) is more significant. At $b_T=2$ GeV$^{-1}$ (which is an optimistic limit of applicability
for the expression (\ref{QCD:F=f})) the deviation is of order $8\%$ for $x=0.1$, whereas at $x=10^{-2}$ it is already $16\%$.

We expect that the next order correction does not bring any significant deviation from the general form of the function $g$ (\ref{QCD:g}). One
can show that the leading large-$b_T$ behavior at $\mathcal{O}(a_s^2)$ order is $\sim \exp(-x^2 b_T^2)x^8 b_T^8 L_T^2$. This implies only some
additional terms to the argument of the logarithm in (\ref{QCD:g}). Such corrections would not significantly influence on the region of the
applicability of the calculation, because it is mostly defined by the exponent factor.

\section{Conclusion}

The TMDs are involved objects which cannot be obtained within the perturbative QCD. For the description of the $b_T$-dependence one should use
some non-perturbative information. In the standard description (see e.g. \cite{Collins:2011zzd}) the $b_T$-dependence of TMDs is given by
unknown function, called non-perturbative factor, which is matched with the integrated PDFs with the help of the first few terms of the
small-$b_T$ OPE.

We suggest the modification of the standard approach, which consists in the consideration of the small-$b_T$ OPE in different operator basis.
While the standard approach uses the power (Taylor-like) expansion, we suggest to consider the expansion which leading term reproduces the
desired form of TMD. Such approach does not spoil the standard properties of TMDs, such as evolution properties. The main prediction of the
approach is the ``fine'' structure upon the general non-perturbative input.

In this article we assume that at middle- and large-$b_T$ the dominating behavior is Gaussian. This is often used in phenomenology conjecture,
(see e.g.\cite{Boer:2011fh,Aidala:2014hva,Aybat:2011zv,Aybat:2011ge}). In order to reproduces this behavior within the leading term of OPE we
uses the operator basis based on Laguerre polynomials (\ref{phi3:Op_Laguerre}). The Laguerre polynomials are also singled out by the fact that
the they are the only classical polynomials with the support $b_T\in(0,\infty)$, which is needed in the case of unpolarized TMD PDFs. We want to
emphasize that there are no theoretical restrictions on the $b_T$ behavior and symmetry properties of TMDs, therefore, basis of another (may be
non-classical) polynomials might describe the experimental data better. The general structure of consideration of TMDs in another basis is the
same as presented in the paper.

The coefficient functions of Laguerre-based expansion have leading behavior $\mathbb{C}_n\sim (b_T/B_T)^n \exp(-b_T^2/4B_T^2)$ (see
(\ref{QCD:Laguerre_Cqq},\ref{QCD:Laguerre_Cqg})). Neglecting the $n>0$ terms of expansion we obtain the perfect Gaussian behavior at the leading
term of perturbative expansion, which is modified by the quantum corrections. In the contrast to the Taylor-like expansion where the corrections
take the form of logarithms, in the Laguerre-based expansion the corrections change the initial exponential behavior and also contain powers and
logarithms of $b_T$. We interpret these corrections as the perturbative modification of the non-perturbative (Gaussian) input. The corrections
reproduce the physical effects which were not incorporated in the initial simple model. For example, the initial Gaussian behavior is
$x$-independent, but the corrections essentially depend on $x$.

In the paper we have presented the detailed calculation of OPE in basis of Laguerre polynomials. The main technical details are given by the
example of the OPE for TMD operator in $\phi^3_{D=6}$ theory. We have calculated the coefficient functions at tree and one-loop accuracy for the
unpolarized quark TMD PDF, the results are presented in equations (\ref{QCD:OPE_Laguerre}-\ref{QCD:Laguerre_Cqg}).

The obtained expression for TMD PDF (\ref{QCD:F=f}) has only two free parameters $b_\text{max}$ and $B_T$. The parameter $b_\text{max}$
describes the matching of small-$b_T$ regime with large-$b_T$ regime and can be restricted by the assumption that the size of the perturbative
correction should be small in comparison to the leading order. The parameter $B_T$ is an unknown parameter of the OPE. In general, the OPE is
independent on $B_T$: it cancels in the sum of all terms. However, restricting our consideration by the first term we introduce artificial
dependence on this scale. Therefore, the parameter $B_T$ is to be found from the comparison with experimental data.

The obtained perturbative results show the general tendency to spread the initial Gaussian input. Moreover the spreading is stronger for smaller
$x$, which is in agreement with the parton picture of hadron. Indeed, the smaller $x$ partons should be distributed in the wider spatial region
in comparison to larger-$x$ partons. At one-loop accuracy, the large-$b_T$ dominating term of coefficient function is proportional to $\exp(-x^2
b_T^2)b_T^4$. One can show that at two loop order the large-$b_T$ dominating term is proportional to $\exp(-x^2 b_T^2)b_T^8$. So, the higher
order contributions can violate the Gaussian behavior suggested by leading term of Laguerre-based OPE.

The influence of the non-perturbative corrections grows with the growing of $b_T$. For large $b_T$ the perturbative contribution to TMD is
negligible, and can only be parameterized by some function. Therefore, the Laguerre-based OPE can be viewed as alternative form of the standard
matching procedure, which allows to reduce the significance of the non-perturbative factor at small $b_T$, but does not neglect it.

\acknowledgments

Author thanks A.Moiseeva, A.Efremov and N.Stefanis for stimulating discussions. The work is supported in parts by the European
Community-Research Infrastructure Integrating Activity Study of Strongly Interacting Matter" (HadronPhysics3, Grant Agreement No. 28 3286) and
the Swedish Research Council grants 621-2011-5080 and 621-2010-3326.

\appendix
\section{Mixing between operators with different number of fields}

In this appendix, we study the mixture of operators within the definition of the OPE, and its influence on the coefficient function calculation
procedure. Since we are interested only in the dimensions of operators and orders of perturbative expansion, we do not label other features of
operators.

The small-$b_T$ OPE is an exact relation of the form
\begin{eqnarray}\label{app:scheme_OPE}
O(x,b_T)=\frac{1}{b_T^2}C_0(x,b_T)\cdot 1+\int_x^1 \frac{dz}{z}\sum_{n=2}^\infty \sum_{k=0}^\infty b_T^{n+k-2} C_{n,k}\(\frac{x}{z},b_T\)
O_{n,k}(z),
\end{eqnarray}
where $O_{n,k}$ is the transversely local operator which contains $k$ transverse derivatives and $n$ fields $\phi$. The functions $C$ are
dimensionsionless, but they contain logarithms of $b_T$. The expansion (\ref{app:scheme_OPE}) is schematic, it counts only the orders of
dimensions.

In order to obtain the coefficient functions $C$ one should consider all possible Green functions of (\ref{app:scheme_OPE}) and match both sides
of the relation. The matching can be done order-by-order in powers of $b_T$ and $g$. With this aim we introduce the notations for the terms of
perturbative expansion. The $N$-field Green function $G_N(b_T)$ of the operator $O(x,b_T)$ has the following pertubative expansion
\begin{eqnarray}\label{app:A2}
G_0(b_T)=\frac{1}{b_T^2}\sum_{a=0,2,..}^\infty g^a G_0^{[a]},~~~G_N=\sum_{k=0}^\infty b_T^k\sum_{a=|N-2|,..}^\infty g^aG_{N,k}^{[a]},
\end{eqnarray}
where summation runs with step 2. The $N$-field Green functions $R_{N,n,k}$ of operators $O_{n,k}$ have the following perturbative expansions
\begin{eqnarray}\label{app:A3}
R_{0,n}=0,~~~R_{N,0}=\sum_{a=N-2}^\infty g^a R_{N,0}^{[a]},~~~ R_{N,n,k}=\sum_{a=|N-n|}^\infty g^a R_{N,n,k}^{[a]},
\end{eqnarray}
where summation runs with step 2. The coefficients of the expressions (\ref{app:A2},\ref{app:A3}) can be obtained by straightforward
calculation. The perturbative expansion of the coefficient functions presumably has all terms of perturbative expansion with unknown
coefficients,
\begin{eqnarray}
C_{n,k}\(x,b_T\)=\sum_{a=0}^\infty g^a C_{n,k}^{[a]}(x,b_T).
\end{eqnarray}

Matching the perturbative expansion on the both sides of (\ref{app:scheme_OPE}) we obtain the expressions for coefficient functions. For
example, at the zeroth order of $b_T$ we have:
\begin{eqnarray}
G_{N,0}^{[a]}=\sum_{b=N-2}^a C^{[a-b]}_{2,0}\ast R^{[b]}_{N,2,0},
\end{eqnarray}
where $a\geqslant N-2$, and $\ast$ represents the Mellin convolution. This system is overcomplete, but the existence of solution is guarantied
by the existence of OPE itself. Therefore, one can choose any complete subset of equations and use it for the evaluation of the coefficients.

Let us find at the relations responsible for the coefficient functions of $n=2$ operators with arbitrary number of derivatives. The coefficient
functions at the leading orders  can be obtained from the equations at $(N,a)=(2,0)$, they read
\begin{eqnarray}\label{app:G^0}
G_{2,k}^{[0]}=C_{2,k}^{[0]}\ast R_{2,2,k}^{[0]}.
\end{eqnarray}
This relation is a simple comparison of the expansion for the tree diagrams. The equations at $(N,a)=(3,0)$, $(2,1)$, and $(4,0)$ imply that
$C_{3,k}^{[0]}=0$, $C_{2,k}^{[1]}=0$, and $C_{4,k}^{[0]}=0$, correspondingly. The equation at $(N,a)=(3,1)$ reads
\begin{eqnarray}\nn
G^{[1]}_{3,k}=C_{2,k}^{[0]}\ast R_{3,2,k}^{[1]}+C_{3,k}^{[1]}\ast R_{3,3,k}^{[0]}.
\end{eqnarray}
However, the diagrams responsible for $G^{[1]}_{3,k}$ do not contain loops, therefore, this relation is fulfilled by the coefficient function
$C_{2,k}^{[0]}$ with $C_{3,k}^{[1]}=0$. The one-loop coefficient functions of operators with $n=2$  are given by the equation at $(N,a)=(2,2)$
\begin{eqnarray}\label{app:G^2}
G_{2,k}^{[2]}=C_{2,k}^{[2]}\ast R_{2,2,k}^{[0]}+C_{2,k}^{[0]}\ast R_{2,2,k}^{[2]}.
\end{eqnarray}
We conclude that one needs to consider  only the 2-point Green function at one-loop in order to obtain $C_{2,k}^{(2)}$. The two-loops expression
for $C_{2,k}$ is already effected  by operators with four fields ($k>1$):
\begin{eqnarray}\label{app:G^4}
G_{2,k}^{[4]}=C_{2,k}^{[4]}\ast R_{2,2,k}^{[0]}+C_{2,k}^{[2]}\ast R_{2,2,k}^{[2]}+C_{2,k}^{[0]}\ast R_{2,2,k}^{[4]}+C_{3,k}^{[3]}\ast
R_{2,3,k}^{[1]}+C_{4,k}^{(2)}\ast R_{2,4,k}^{[2]}.
\end{eqnarray}
The higher-loop orders would include the operators with greater number of fields.

This exercise shows us that the coefficient function for the operators with two-field cannot be defined without simultaneous definition of all
other operators. Nonetheless, the situation is not hopeless since the mixture between the operators increases order-by-order. It means that the
coefficients $C^{[2]}$ can be obtained by considering the two-fields operators only. While the coefficients $C^{[4]}$ require the consideration
of two- and four-fields operators.

\section{Expression for the diagrams in QCD}
In this section we collect the expressions needed for the calculation of coefficient functions. The calculations are performed in the Feynman
gauge within the dimension regularization (for the regularization of ultraviolet divergences) and $\delta$-regularization (for the
regularization of the rapidity divergences). Since the calculation is done for the general external momentum we do not need any additional
infrared regularization.

The $\delta$-regularization is introduced in the fashion which resembles the regularization used in the SCET-calculation
\cite{GarciaEchevarria:2011rb,Chiu:2009yx}. Namely, the radiation of a gluon with momentum $p$ by a half-infinite Wilson line (say, pointing in
the direction $n^\mu$) results to the integrals of the form
\begin{eqnarray}
\int d^4z\int_0^\infty  ~d\sigma n^\mu \delta^{(4)}(n^\mu \sigma -z)e^{\pm ikz}=\lim_{\tau\to\infty} \frac{\pm in^\mu}{k^+}\(e^{\pm i
k^+\tau}-1\).
\end{eqnarray}
These integrals result to some generalized functions, which can be equivalently rewritten
\begin{eqnarray}
\lim_{\tau\to\infty} \frac{1-e^{\pm i k^+\tau}}{k^+}~~\longleftrightarrow ~~\lim_{\delta^+\to0} \frac{1}{k^+\pm i\delta^+}.
\end{eqnarray}
The superscript ``plus''(``minus'') marks the $\delta$-regulator for the eikonal propagator of the Wilson line in the $n$-($\bar n$)direction.
Keeping the explicit $\delta$-dependence we regularize the rapidity divergences.

\begin{figure}[t]
\begin{center}
\includegraphics[width=0.5\textwidth]{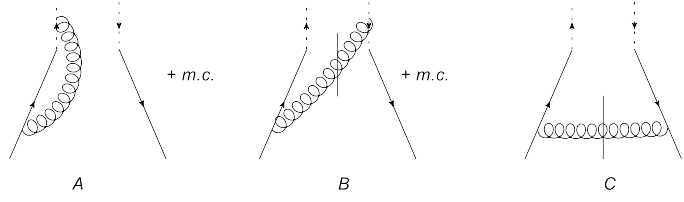}
\caption{One-loop graph describing the leading correction to the TMD unsubtracted operator in QCD. The abbreviation \textit{m.c.} denotes
mirror-conjugated diagrams. The cut propagators denote the Schwinger propagators. } \label{fig:QCD_QQ_graphs}
\end{center}
\end{figure}

The tree order of the Green function (\ref{QCD:G2_tree}) reads
\begin{eqnarray}
\mathcal{G}_{q/q}^{[0]}(x,b_T,p)=\delta(1-x)e^{-i(bp)_T}\frac{\gamma^+}{2p^+}.
\end{eqnarray}

The $\alpha_s$-correction to the Green function is given by a number of contribution. The corrections to the unsubtracted operator are given by
the diagrams shown in fig.\ref{fig:QCD_QQ_graphs}. In the limit $\epsilon\to 0$ and $\delta\to 0$ the contribution of individual diagrams reads
\begin{eqnarray}
G^{[2]}_{A}&=& a_sC_F\frac{\gamma^+}{p^+}\delta(1-x)e^{-i(bp)_T}e^{\epsilon\gamma_E}\Gamma(\epsilon)
\Big[2\(\frac{\mu^2}{-p^2}\)^\epsilon\(1+\ln\(\frac{\delta^+}{p^+}\)\)\\\nn &&~~~~~~~~~~~~~~~~~~~~~~~~~~~~~~~~~-
\epsilon\(-4+\frac{5\pi^2}{12}+\ln^2\(\frac{\delta^+}{p^+}\)\) +\mathcal{O}(\epsilon)+\mathcal{O}(\delta^+)\Big],
\\
G^{[2]}_{B}&=&a_sC_F\frac{\gamma^+}{p^+}e^{-ix(bp)_T}e^{\epsilon\gamma_E} \Big[
\(\frac{2x}{(1-x)_+}-2\delta(1-x)\ln\(\frac{\delta^+}{p^+}\)+\mathcal{O}(\delta^+)\)\bar f_\epsilon \\\nn&& ~~~~~~~~~~~~~~~~~~~~~~~~~~~~~~~~~
-\Gamma(1+\epsilon) \(\frac{\mu^2}{- p^2}\)^\epsilon\delta(1-x)\(2-\frac{\pi^2}{12}-\ln^2\frac{\delta^+}{p^+}\)\Big],
\\
G^{[2]}_{C}&=&a_s C_F (1-\epsilon)\bar x e^{-ix(bp)_T}e^{\epsilon\gamma_E} \Big(-2x^2p^+\gamma^-f_{\epsilon+1}
-2x^2\not{\!p}_Tf_{\epsilon+1}-ix\not{\! b}_T f_\epsilon
\\\nn&&~~~~~~~~~~~~~~~~~~~~~~~~~~~~~~~~~~~~
 +\frac{\gamma^+}{p^+}\(f_\epsilon-x^2p_T^2f_{\epsilon+1}-ix(bp)_Tf_\epsilon -x\bar x p^2f_{\epsilon+1}\)\Big),
\end{eqnarray}
where the limit $\delta^+\to0$ has been taken in the sense of distribution, the factor $e^{\epsilon\gamma_E}$ comes from the $\overline{MS}$
scheme, and $a_s=\frac{g^2}{(4\pi)^2}=\frac{\alpha_s}{4\pi}$. The function $f_a$ is
\begin{eqnarray}
f_a=f_a(x,b,p)&=&2e^{-i\pi \epsilon}\mu^{2\epsilon}\(\frac{-b_T^2-i0}{4x\bar x p^2+i0}\)^\frac{a}{2}K_a\(\sqrt{(x\bar x
p^2+i0)(-b_T^2-i0)}\)\\\nn &=&i\pi e^{-i\pi \epsilon}\mu^{2\epsilon}\(\frac{b_T^2}{4x\bar x p^2}\)^\frac{a}{2}H_a^{(1)}\(\sqrt{x\bar x
p^2b_T^2}\),
\end{eqnarray}
where for the second equality we have assumed that $p^2>0$. The function $\bar f_\epsilon$ is a shorthand notation for the regular at $x\to 1$
part of $f_\epsilon$
$$
\bar f_\epsilon=f_\epsilon+\Gamma(\epsilon)\(\frac{\mu^2}{-xp^2}\)^\epsilon(1-\bar x^{-\epsilon}).
$$

The quark-field renormalization gives the contribution
\begin{eqnarray}
Z^{[2]}_q(\mu)\mathcal{G}_{q/q}^{[0]}(b,p)&=& -a_sC_F\frac{\gamma^+}{p^+}\delta(1-x)e^{-i(bp)_T}e^{\epsilon\gamma_E}
\(\frac{\mu^2}{-p^2}\)^{\epsilon}\Gamma(\epsilon)\frac{\Gamma^2(2-\epsilon)}{\Gamma(3-2\epsilon)}.
\end{eqnarray}

\begin{figure}[t]
\begin{center}
\includegraphics[width=0.35\textwidth]{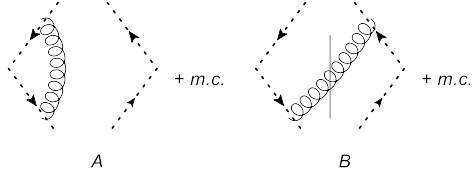}
\caption{One-loop graph describing the leading correction to the soft factor. The abbreviation \textit{m.c.} denotes mirror-conjugated diagrams.
The cut propagators denote the Schwinger propagators. } \label{fig:QCD_SF_graphs}
\end{center}
\end{figure}

The contribution of the soft factor is represented by diagrams in fig.\ref{fig:QCD_SF_graphs}. Their expression reads
\begin{eqnarray}
S^{[2]}_A&=& -4a_sC_Fe^{\epsilon\gamma_E}\(\frac{\mu^2}{2\delta^+\delta^-}\)^{\epsilon} \Gamma^2(\epsilon)\Gamma(1-\epsilon),
\\\label{app:S2B}
S^{[2]}_B&=&-4a_sC_Fe^{\epsilon\gamma_E}\Big[\Gamma(-\epsilon)\(\ln\(2\delta^+\delta^-\frac{b_T^2}{4}\)+\gamma_E-\psi(-\epsilon)\)\(\frac{\mu^2b_T^2}{4}\)^\epsilon
\\&&\nn~~~~~~~~~~~~~~~~~~~~~~~~~~~~~~~~~~~~-
\(\frac{\mu^2}{2\delta^+\delta^-}\)^\epsilon\Gamma(\epsilon)\pi\text{ctg}(\pi \epsilon)+\mathcal{O}(\delta)\Big].
\end{eqnarray}

The obtained results in the limit $b_T^2\to 0$ can be compared with the results presented in \cite{GarciaEchevarria:2011rb}. They are in
agreement up to a finite constant, which is result of different infrared regularization. Also we note the difference by factor two (in
comparison with corresponded expression in \cite{GarciaEchevarria:2011rb}) for the argument of logarithms of $\delta^+\delta^-$ in
(\ref{app:S2B}). This is a result of different normalization of light-cone vectors (we use normalization $(n\bar n)=1$, whereas in
\cite{GarciaEchevarria:2011rb} $(n\bar n)=2$ is used).

The expression for the second order of the Green function reads
\begin{eqnarray}
\mathcal{G}_{q/q}^{[2]}(x,b_T,p)&=&G^{[2]}_A+G^{[2]}_B+G^{[2]}_C+Z^{[2]}_q\mathcal{G}_{q/q}^{[0]}-\frac{1}{2}\mathcal{G}_{q/q}^{[0]}\(S^{[2]}_A+S^{[2]}_B\).
\end{eqnarray}
The parameters $\delta^\pm$ should be taken according to prescription (\ref{QCD:deltapm}). In the limit $p^2\to 0$ this expression reads
\begin{eqnarray}
\mathcal{G}_{q/q}^{[2]}(x,b_T,p)&=&a_s C_F \frac{\gamma^+}{2p^+}e^{-ix(bp)_T}e^{\epsilon \gamma_E}\Bigg\{ \\&&\nn
 \delta(1-x)\Bigg[2\Gamma(-\epsilon)\(\frac{\mu^2b_T^2}{4}\)^\epsilon\(\ln\(\frac{b_T^2 \zeta}{4}\)+\gamma_E-\psi(-\epsilon)\)
 \\\nn &&
 +\(\frac{\mu^2}{-p^2}\)^\epsilon\Gamma(\epsilon)\(4-2\frac{\Gamma^2(2-\epsilon)}{\Gamma(3-2\epsilon)}\)
 \\\nn&&
 +\(\frac{\mu^2}{-p^2}\)^\epsilon\Gamma(1+\epsilon)\(-4+\frac{\pi^2}{6}+2\ln^2\(\frac{\delta^+}{p^+}\)\)
 \\\nn&& \qquad\qquad\qquad\qquad+\(8+\frac{\pi^2}{6}-2\ln^2\(\frac{\delta^+}{p^+}\)+\mathcal{O}(\epsilon)\)\Bigg]
 \\ \nn&&+\(\frac{\mu^2}{-x\bar x p^2}\)^\epsilon\Gamma(\epsilon)\Big(2\bar x(1-\epsilon)(1-ix(bp)_T) -2\bar x \epsilon(1-\epsilon)
 \\\nn&&\qquad\qquad\qquad\qquad\qquad\qquad\qquad-2x\frac{p_T^2}{p^2}\epsilon(1-\epsilon)- x^2\bar x (1-\epsilon)\frac{b_T^2p_T^2}{4} \Big)
 \\ \nn&&
 +\(\frac{\mu^2}{-xp^2}\)^\epsilon\Gamma(\epsilon)\frac{4x}{(1-x)_+}
 \\ \nn &&+\(\frac{b_T^2\mu^2}{4}\)^\epsilon\Gamma(-\epsilon)\(\frac{4x}{(1-x)_+}+2\bar x (1-\epsilon)(1-ix(b p)_T)
 -\frac{\bar x x^2}{4}b_T^2p_T^2\frac{1-\epsilon}{1+\epsilon}\),
\end{eqnarray}
where we have also eliminated the terms proportional to $\gamma_T$ and $\gamma^-$.  One can see that this expression is independent on $\delta$
at $\mathcal{O}(\epsilon)$.

The term $\sim p_T^2/p^2$, although it does not contribute to any coefficient function at $\mathcal{O}(a_s)$, implies the necessity to introduce
the operator $\partial_T^2$ (without any fields) into the OPE. The coefficient function of this operator is non-zero with leading term $\sim
a_s^2$.

\begin{figure}[t]
\begin{center}
\includegraphics[width=0.16\textwidth]{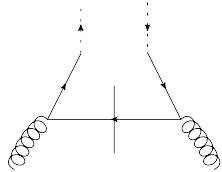}
\caption{One-loop graph describing the leading contribution to the TMD operator with the gluon external fields. The cut propagator denotes the
Schwinger propagator. } \label{fig:QCD_QG_graphs}
\end{center}
\end{figure}

The Green function with gluon field at $\mathcal{O}(g^2)$ is described by a single diagram shown in fig.\ref{fig:QCD_QG_graphs}. Due to the
absence of tree-order diagram there is no contribution from the soft factor and field renormalization constants. Since we are interested only in
the projection of leading twist operators, we average the gluon states over polarizations. In this way the diagrams with Wilson-line --
external-gluon interaction vanish. Evaluating the diagram we obtain
\begin{eqnarray}
\mathcal{G}_{q/g}^{[2]}(x,b_T,p)=4a_se^{-ix(bp)_T}e^{\epsilon\gamma_E}\[(1-\epsilon-2x\bar x)f_\epsilon-x\bar x(1-\epsilon)p^2
f_{\epsilon+1}\](1-\epsilon)^{-1},
\end{eqnarray}
where the factor $(1-\epsilon)^{-1}$ is the result of the gluon-polarization averaging. In the limit $p^2\to0$ this expression reads
\begin{eqnarray}&&
\mathcal{G}_{q/g}^{[2]}(x,b_T,p)=4a_se^{-ix(bp)_T}e^{\epsilon\gamma_E}\Bigg[\\&&\nn\qquad\(1-\frac{2x\bar
x}{1-\epsilon}\)\(\Gamma(\epsilon)\(\frac{\mu^2}{-x\bar x p^2}\)^\epsilon+\Gamma(-\epsilon)\(\frac{\mu^2b_T^2}{4}\)^\epsilon\)-
\Gamma(1+\epsilon)\(\frac{\mu^2}{-x\bar xp^2}\)^{\epsilon}\Bigg].
\end{eqnarray}
This result in the limit $b_T\to0$ coincides with the result of calculation performed in \cite{Aybat:2011zv}.

\end{document}